\documentclass[a4paper,11pt]{article}

\usepackage{graphicx}
\usepackage{float}
\usepackage{wrapfig}
\usepackage{xcolor}
\usepackage[T1]{fontenc}
\usepackage{epsfig}
\usepackage{color}
\usepackage{amsmath,jheppub,mathtools}
\usepackage{subfloat}
\usepackage{amsfonts}
\usepackage{braket}
\usepackage{cleveref}
\usepackage{epstopdf}
\usepackage{caption}
\usepackage{subcaption}
\usepackage{enumitem}
\usepackage[numbers]{natbib}
\usepackage[titletoc,toc,title]{appendix}
\usepackage{hyperref}
\hypersetup{
	colorlinks=true,
	linkcolor=blue,
	filecolor=red,      
	urlcolor=blue,
	citecolor=blue
} 

\title{\boldmath Notes on heating phase dynamics in Floquet CFTs and Modular quantization}

\author[a]{Suchetan Das}
\author[b]{Bobby Ezhuthachan,}
\author[b]{Somnath Porey,}
\author[b]{Baishali Roy}
\affiliation[a]{Center for High Energy Physics,
Indian Institute of Science, Bengaluru 560012, India}
\affiliation[b]{Ramakrishna Mission Vivekananda Educational and Research Institute, Belur Math, Howrah-711202, West Bengal, India}

\emailAdd{suchetan1993@gmail.com}
\emailAdd{bobby.phy@gm.rkmvu.ac.in}
\emailAdd{somnathhimu00@gm.rkmvu.ac.in}
\emailAdd{baishali.roy025@gm.rkmvu.ac.in}

\date{}

\abstract{In this article, we explore the connection between the heating phase of periodically driven CFTs and the Modular Hamiltonian of a subregion in the vacuum state. We show that the heating phase Hamiltonian corresponds to the Modular Hamiltonian, with the fixed points mapping to the endpoints of the subregion. In the bulk dual, we find that these fixed points correspond to the Ryu-Takayanagi surface of the AdS-Rindler wedge. Consequently, the entanglement entropy associated to the boundary interval within two fixed points exactly matches with the Rindler entropy of AdS-Rindler. We observe the emergent Virasoro algebra in the boundary quantization of the Modular Hamiltonian has a striking similarity with the emergent near Horizon Virasoro algebra. This is a consequence of the fact that while obtaining the boundary Virasoro algebra, a cut-off with conformal boundary condition around the fixed point is introduced, which in the bulk is related to a stretched horizon, with an emergent two-dimensional conformal symmetry. We also argue that as one tunes the parameter space of Floquet Hamiltonians to transition from the non-heating to the heating phase the operator algebra type changes from Von Neumann type $I$ to $III_1$ factor, providing a non-equilibrium analogue of the Hawking-Page transition. 

}

\begin{document} 
\maketitle
	
	\flushbottom


			\section{Introduction}

The study of Floquet CFTs has recently garnered considerable attention as analytically tractable models of physics out of equilibrium.  At stroboscopic times, the evolution of such systems is governed by an effective Hamiltonian, referred to in the literature as the Floquet Hamiltonian $H_F$. One of the novel features of their dynamics is the existence of three different dynamical phases, characterized by qualitatively different temporal growths of the entanglement entropy and energy. In particular, one of the phases, aptly referred to as the {\it heating phase}, is characterized by an exponentially growing energy and linearly growing entanglement entropy in stroboscopic time. Moreover, for the special example of a $sl(2)$ Floquet CFT, an OTOC computation in large-$c$ CFTs in the heating phase also shows a Lyapunov growth, characteristic of early time chaos \cite{Das:2022jrr}. In the bulk description, such a Lyapunov growth in a holographic CFT is understood to be a consequence of blue-shifting of energy in the presence of a horizon \cite{Shenker:2013pqa}. Indeed, as was confirmed in \cite{Das:2022pez}, there is an observer-dependent horizon in the dual description of the $sl(2)$ Floquet CFT in the heating phase \footnote{For other works on the holographic bulk dual of Floquet CFTs or deformed CFTs see \cite{deBoer:2023lrd, Jiang:2024hgt, Kudler-Flam:2023ahk, MacCormack:2018rwq, Caputa:2022zsr}.}. As was shown in that paper, one can reproduce the Lyapunov exponent from a bulk computation, following the methods of \cite{Shenker:2013pqa}. 

These features of the heating phase are best explained by noting that $H_F$ is also the Modular Hamiltonian of a line interval on the circle in the vacuum state of the CFT, with the endpoints of the interval being the two fixed points of the flow under $H_F$\footnote{This point has been made previously, for e.g.- see\cite{Khetrapal:2022dzy}}, as we show in section [\ref{mhhp}]. Thus, the Floquet CFT in the heating phase captures the dynamics of this sub-system in modular time. Consequently, in the bulk, it describes the dynamics within the corresponding Entanglement Wedge (EW). In fact, as we show in section [\ref{fprt}], in the bulk, the fixed points of the flow under $H_F$ correspond to the RT surface. 


In \cite{Das:2022pez}, the bulk dual of the heating phase was described in the frame of an observer moving under the flow generated by the bulk representation of $H_F$, (which we refer to as $H^b_F$). In the boundary theory, one can analogously study the heating phase in the CFT directly in a conformal frame\footnote{We call this frame in which time is the modular time, i.e. generated by $H_F$ as $CF_2$, while the frame, where the time is the one generated by the undeformed Hamiltonian is referred to as $CF_1$.} where the modular time\footnote{This is the same as the stroboscopic time of the driven CFT} is the real-time of the system. 
Quantization in this frame would provide the boundary description of the bulk dynamics inside the Entanglement Wedge. As has been emphasized before in the literature \cite{Ohmori:2014eia}, to make this quantization well-defined, one has to impose boundary conditions around the two endpoints (fixed points of the Floquet Hamiltonian)\footnote{after imposing a cut-off ($\epsilon$) around the fixed points}. The full Hilbert space of the theory can formally be then expressed as, 
\begin{equation}\label{RQG}
\mathcal{H}_{CFT} = \bigoplus_{B_1, B_2} \mathcal{H}_{B_1,B_2}^{u}\otimes\mathcal{H}_{B_2,B_1}^{d}.     
\end{equation}
    
\begin{figure}
\centering
\includegraphics[scale= 0.7]{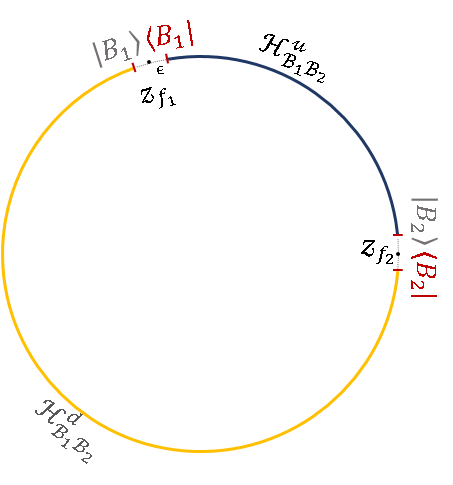}
\caption{ The illustration depicts a constant time slice, with the blue arc representing the subregion $u$ between two fixed points denoted by black dots. Similarly, the yellow arc corresponds to its complement subregion $d$. The Hilbert spaces associated with each are denoted as $\mathcal{H}^u_{B_1B_2}$ and $\mathcal{H}^d_{B_1B_2}$. $B_1$ and $B_2$ represent the boundary conditions applied at the two ends of $u(d)$ at a distance $\epsilon$ from the fixed points $z_{f_1}$ and $z_{f_2}$.}
\label{fig1}
\end{figure}

Where $B_1$, $B_2$ represent the boundary conditions imposed at a cut-off distance $\epsilon$ away from the two fixed points \cite{Ohmori:2014eia, Agia:2022srj}. Here, $\mathcal{H}^{u(d)}_{B_1, B_2}$, is dual to the region inside the Entanglement Wedge, while the $\mathcal{H}^{d(u)}_{B_1, B_2}$, would be dual to the complimentary Wedge (see Fig.\ref{fig1}). If we impose conformal boundary conditions at the cut-off, then the boundary condition will be preserved by one copy of the Virasoro algebra. Usual AdS/CFT rules imply that the bulk description must also involve a boundary placed at a small cut-off away from the RT surface. This is very reminiscent of the stretched Horizon scenario for black holes \cite{tHooft:1984kcu, Susskind:1993if}. Indeed, as we show in section [\ref{bulk}], if one pulls the conformal transformation from $CF_1$ to $CF_2$ into the bulk \cite{Anand:2017dav}, we get the AdS-Rindler metric. In fact, in these coordinates, the RT surface maps to the Rindler Horizon. Moreover, as we argue in section [\ref{bulk vir}], the `emergent' boundary Virasoro algebra in the $CF_2$ frame, should be identified with the near horizon Virasoro algebra that has been discussed previously in the literature \cite{Solodukhin:1998tc, Carlip:2002be, Carlip:1999db, Carlip:2005zn, Guica:2008mu, Kang:2004js}.

In \cite{Das:2022pez}, it was shown that the bulk dual metric of the heating phase is given by an $AdS_2$ black hole patch of $AdS_3$. As we show in section [\ref{bulk}], this metric and the AdS Rindler metric are related by a diffeomorphism in the bulk and represent the same bulk observer\footnote{Consequently, they are not large diffeomorphisms.}. 
Since the Modular Hamiltonian, generates evolution along the closed lines around the fixed points, (in the Euclidean plane), the entanglement entropy of the line interval is by definition equal to the thermal entropy of the associated (Euclidean) cylinder (see Fig[\ref{fig1}])\footnote{This should match the AdS Rindler entropy in the bulk}.
While this entropy would depend on the boundary condition imposed, the leading contribution (in the $\epsilon\rightarrow 0$) is independent of it \cite{Cardy:2016fqc}. In the full tensor product Hilbert space description given by equation [\ref{RQG}], the purified thermofield double state should be identified with the vacuum state of the CFT $|0\rangle$:
\begin{equation}
    |0\rangle \equiv \sum_{B_1, B_2} \sum_{\Delta_{B_1, B_2}} e^{-\frac{\Delta_{B_1, B_2}}{2}}|\Delta_{B_1, B_2}\rangle_{(u)}\otimes |\Delta_{B_2,B_1}\rangle_{(d)}
\end{equation}
Where $\Delta_{B_1, B_2}$ are the spectrum of $H^{(u)}_{B_1, B_2}$ and should be identified in the bulk with the microstates of the AdS Rindler Geometry \cite{Czech:2012be}\footnote{ Formally, the Modular Hamiltonian of the region $l$, is given by $K \equiv H^{(u)}_{B_1, B_2}\otimes {\bf I}^{(d)}_{B_2, B_1} - {\bf I}^{(u)}_{B_1, B_2}\otimes H^{(d)}_{B_2, B_1}$}. The States $|\Delta_{B_1, B_2}\rangle_{(u)}$ are the eigenstates of $H^{(u)}_{B_1, B_2}$, and in the bulk will be dual to the fixed area states \cite{Dong:2018seb}.



The rest of the paper is organized as follows: In the next section, after a brief review of $sl(2,\mathbb{R})$ driven CFTs in section [\ref{dfcft}], we show the equivalence of the heating phase Hamiltonian with the Modular Hamiltonian of a sub-region in the CFT vacuum in section [\ref{mhhp}]. We end this section by discussing the classification of the dynamical phases in terms of Von Neumann factors.  We explicitly show in section [\ref{eqv}], that the metric inside the Entanglement Wedge is given by the AdS Rindler metric, and that the Rindler Horizon is the RT surface. 
In section [\ref{fprt}], we show that in the bulk, the fixed points of the flow correspond to the Ryu-Takayanagi surface, which ends on the boundary fixed points. In Section [\ref{hpq}], we provide a brief review of heating phase quantization in the conformal frame $CF_2$ and the corresponding Virasoro algebra, as discussed in \cite{Tada:2019rls}. In Section [\ref{bulkvir}], we begin by reviewing the work of \cite{Solodukhin:1998tc}, where the near-horizon Virasoro algebra was constructed. We demonstrate that the structure of the central extensions in both cases matches exactly, thus providing evidence for the identification of the emergent Virasoro algebra in the $CF_2$ frame with the near-horizon algebra in the bulk.


In section [\ref{EE}], we explicitly show that the entanglement entropy of the boundary CFT is indeed the AdS Rindler entropy. We end with a summary of our results and a discussion of possible future direction in section [\ref{disc}]. The article has two appendices. In Appendix \ref{apnA}, we show that the non-heating phase Hamiltonian is related to the scaling generators by a unitary transformation, while in Appendix \ref{apnB}, some of the details of section [\ref{mhhp}] is presented. 



\section{Floquet Hamiltonian as a Modular Hamiltonian}
To make the article self-contained, we will first briefly review the dynamics of a driven CFT, where the Hamiltonian of the system is taken to vary periodically in time, with period T. Depending on the chosen protocol, within each time period (T), the Hamiltonian may be changing continuously or in discrete steps. For the sake of simplicity, we focus on the discrete-drive protocol here.

 \subsection{Brief review of driven Floquet CFTs}\label{dfcft}
 We consider a system that is defined on a ring of length $L$. The general form of the Hamiltonian  ($H_{def}^{(i)}$) at the $i^{th}$ discrete step, which lasts for a time $T_i$, is :
 \begin{equation}
		H_{def}^{(i)} = \int^L_0 f_{i}(x) T_{00} (x)  dx
	\end{equation}
The functions $f_i(x)$ and time interval $T_i$ are part of the given protocol. Restricting to $sl(2, \mathbb{R})$ drives, we choose the functions such that the Hamiltonians are built out of global generators. With this restriction, the general form of `$f_i(x)$' becomes:
\begin{equation*}
    f_i(x)=(a_0^{(i)}+ b_0^{(i)} \cos{\frac{2 \pi x}{L}}+ c_0^{(i)} \sin{\frac{2 \pi x}{L}})
\end{equation*}

The corresponding Hamiltonian, can schematically be represented as:

\begin{equation*}
		H=\sum_{n=0}^{\pm 1} p_n L_n+\sum_{n=0}^{\pm 1} q_n\overline{L}_n- \frac{c}{12}
	\end{equation*}

The coefficients $p_n$ and $q_n$ depends on the specific choice of $a_0^{(i)}, b_0^{(i)}$ and $c_0^{(i)}$.  For instance, the choice $(a_0^{(i)}=1, b_0^{(i)}=c_0^{(i)}=0)$, corresponds to the usual CFT Hamiltonian $H_0=(L_0+ \Bar{L}_0)$, while $(a_0^{(i)}=1, b_0^{(i)}=-1, c_0^{(i)}=0)$, corresponds to the so-called sine-squared deformed Hamiltonian $H_1=\left(L_0+\left( \frac{L_1+L_{-1}}{2}\right)+ c.c.\right)$\footnote{$L_n$ are the Virasoro generators, defined in the usual way- $L_{n}=\frac{c}{24}\delta_{n,0}+\frac{L}{2 \pi}\int_{0}^{L}\frac{dx}{2 \pi}e^{\frac{2 \pi i n x}{L}}T(x)$ and $\bar{L}_{n}=\frac{c}{24}\delta_{n,0}+\frac{L}{2 \pi}\int_{0}^{L}\frac{dx}{2 \pi}e^{-\frac{2 \pi i n x}{L}}\bar{T}(x)$. Here, $T(x)= \pi \left(T_{00}(x)+T_{01}(x) \right)$ and $\bar{T}(x)= \pi \left(T_{00}(x)-T_{01}(x) \right)$.}.
 

As an example of a simple drive protocol, consider evolving the system for time $T_0$
with the Hamiltonian $H_0$ and then change the Hamiltonian to $H_1$ and evolve it for $T_1$ and keep repeating this process, so that the time period is $T =T_0 + T_1$. At stroboscopic time (n$T$), the dynamics of the system are governed by an effective (Floquet) Hamiltonian. For our choices of the $f_i(x)$, the Floquet Hamiltonian of the system will always be some linear combination of the global Virasoro modes since they form a closed algebra. The form of the Floquet Hamiltonian can be derived by the requirement that
\begin{equation*}
    \left( e^{i H_1 T_1} e^{i H_0 T_0}\right)^n \mathcal{O}(z_i, \Bar{z}_i) \left(e^{-i H_0 T_0} e^{-i H_1 T_1}\right)^n \equiv e^{i n H_{F} T} \mathcal{O}(z_i, \Bar{z}_i) e^{-i n H_{F} T}
\end{equation*}

For any local operator $\mathcal{O}$. For the choices of driving Hamiltonian, that we are considering here, the most general form of the Floquet Hamiltonian is,

\begin{equation}\label{hdefor}
    H_{F}= (\alpha L_0 +\beta L_{1}+\gamma L_{-1})+ c.c -\frac{c}{12}
\end{equation}

The above Hamiltonian leads to three distinct dynamics, depending on the quantity $(\delta=\alpha^2-4 \beta \gamma)$ defined using the coefficients $\alpha,\ \beta,\ \gamma$. The three classes of Hamiltonians for $\delta\  (=0,\ <0,\ >0)$ are known respectively as parabolic, hyperbolic, and elliptic in the literature. It has been shown that two-point and one-point correlation functions e.g. entanglement entropy and total energy show different dynamical behaviour when evolved with these three types of Hamiltonians. The entanglement entropy shows linear, oscillatory, and logarithmic behavior whereas the total energy shows exponential, oscillatory, and linear behaviour for $\delta<0,>0$, and $\delta=0$ respectively. Following the authors in \cite{Wen:2018agb, Wen:2020wee, Fan:2020orx, Wen:2022pyj} we will be referring to the three classes as heating phase $(\delta<0)$ and non-heating phase $(\delta>0)$ separated by a phase boundary $(\delta=0)$. 

In the next section, we will show that for the case $\delta <0$, $H_F$ can be expressed as the Modular Hamiltonian of a subregion. 
	

\subsection{Heating phase as Modular Hamiltonian}\label{mhhp}
 
 The Modular Hamiltonian\footnote{By which we mean extended Modular hamiltonian} for a sub-region ($R_1$, $R_2$) in the vacuum of a CFT, defined on a ring of length L is \cite{Cardy:2016fqc},
\begin{equation}\label{modu}
    K= \frac{L}{\pi}\int_{0}^{L}\frac{\sin{\frac{\pi (x-R_1)}{L}}\sin{\frac{\pi(R_2-x)}{L}}}{\sin{\frac{\pi (R_2-R_1)}{L}}} T_{00}(x) dx
\end{equation}
If we compare \eqref{hdefor} and \eqref{modu} to find out the subregion for which  \eqref{hdefor} corresponds to a Modular Hamiltonian, we get
\begin{eqnarray}
    \alpha &=& \cot{\frac{\pi(R_1-R_2)}{L}}\\\label{modular11}
    (\beta+\gamma) &=&-\frac{\cos{\frac{\pi(R_1+R_2)}{L}}}{\sin{\frac{\pi (R_1-R_2)}{L}}}\\\label{modular12}
    i(\beta-\gamma) &=&- \frac{\sin{\frac{\pi(R_1+R_2)}{L}}}{\sin{\frac{\pi (R_1-R_2)}{L}}}\label{modular13}
\end{eqnarray}
Therefore, the quadratic Casimir is,
\begin{equation*}
    \delta=\alpha^2 - 4\beta \gamma =-1
\end{equation*}
Since the quadratic Casimir is negative,  we can conclude that the Hamiltonian in the heating phase corresponds to the extended Modular Hamiltonian of an interval in the vacuum state of a CFT, defined on a ring of length L \footnote{This point was previously mentioned in \cite{Lapierre:2019rwj}.}. In \cite{Tada:2019rls}, the authors considered a simpler choice where $\beta=\gamma$. In that case, 
\begin{eqnarray}\label{modular}
     \alpha=-\cot{\frac{2\pi R}{L}}\\\nonumber
     \beta = \frac{1}{2}\csc \frac{2\pi R}{L}
\end{eqnarray}
In the above case also we can come to the same conclusion since the Casimir is again 
\begin{equation*}
    \delta=\alpha^2 - 4\beta^2 =-1
\end{equation*}
As described in \cite{Tada:2019rls}, when we quantize (\ref{hdefor}) for $(\delta<0)$, the constant time $t$ curve will end at the fixed point of the quantization which is located at $z=\mathcal{Z}_{f_{1,2}} = \frac{-\alpha\pm \sqrt{\alpha^{2}-4\beta^{2}}}{2\beta}$. Using (\ref{modular}), we can see that $\mathcal{Z}_{f_{1,2}} = e^{\pm \frac{2\pi i R}{L}}$. Therefore, these two fixed points correspond to two endpoints of the interval $(R,-R)$. The more general case, when ($\beta \neq \gamma$), is discussed in detail in appendix \ref{apnB}. 

Thus, the Floquet dynamics in the heating phase describes the evolution of operators in modular time\footnote{now identified with the stroboscopic time}, inside the corresponding Entanglement Wedge. This, {\it Modular dynamics}, can be better studied by going to a conformal frame ($CF_2$), where the modular time is identified as time. The conformal map to this new frame is given by:

    \begin{eqnarray}\label{bdyconftran}
    \omega &=& \tau+i x = \frac{\sqrt{|\delta|}}{2 \beta}\tan\left[\frac{\sqrt{|\delta|} z}{2}\right]\\\nonumber
    \Bar{\omega} &=& \tau-i x = \frac{\sqrt{|\delta|}}{2 \beta}\tan\left[\frac{\sqrt{|\delta|}\Bar{z}}{2}\right]
\end{eqnarray}

\begin{figure}
    \centering
    \includegraphics[scale=0.8]{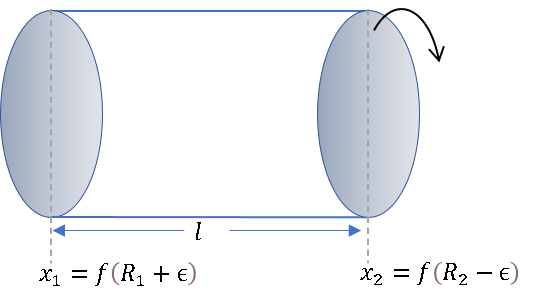}
    \caption{The reduced density matrix for a single interval subregion with endpoints at $R_1$ and $R_2$ can be calculated as a path integral on a Euclidean spacetime $z$, which can be conformally mapped via $w=f(z)$ to a cylinder of fixed width $l=f(R_2-\epsilon)-f(R_1+\epsilon)$ after removing circular regions of radius $\epsilon$ from the two ends of the subregion. The corresponding modular Hamiltonian $K$, which creates flow along the entangling points at the ends of this subregion, is defined on a horizontal line of this cylinder, generating flow along the circular direction.}
    \label{figcylinder}
\end{figure}

\noindent The holomorphic coordinates ($\omega$) describe a finite size cylinder of length $l= 2\log\left[\frac{L \sin \frac{\pi (R_2-R_1)}{L}}{\pi \epsilon}\right]$ as shown in Fig. \ref{figcylinder}. 

To describe a consistent theory on the sub-region, one needs to impose boundary conditions on the fixed points, after cutting off a small region ($\epsilon$) around it\cite{Ohmori:2014eia}. Imposing conformal boundary conditions, one ends up with a BCFT on the cylinder, with the Modular Hamiltonian generating translations along the Euclidean {\it modular time} circle \footnote{since the Modular Hamiltonian generates rotations around the fixed points \cite{Cardy:2016fqc}).}.

By definition, the BCFT on the Euclidean cylinder represents a thermal state, whose thermal entropy is the same as the entanglement entropy of the interval. This construction has been used to compute the entanglement entropy in 2D CFTs \cite{Holzhey:1994we}. The spectrum of the Modular Hamiltonian would then be the thermal spectrum of a BCFT of length $l= 2\log\left[\frac{L \sin \frac{\pi (R_2-R_1)}{L}}{\pi \epsilon}\right]$. In the limit $\epsilon\rightarrow 0$, this becomes effectively an infinite length system and therefore one expects that this spectrum to become continuous in this limit \footnote{This provides us with an effective regulator for 2D CFTs on an infinite line \cite{Tada:2019rls}.  See \cite{Das:2024mlx}, for a detailed analysis of the spectrum in this Quantization.}.

This should be contrasted with the non-heating phase, where the spectrum is discrete \cite{Ishibashi:2016bey}. Moreover, as we show in appendix \ref{apnA}, the Hamiltonian in the non-heating phase is unitarily equivalent to the generator of scaling transformations. Therefore, the non-heating phase dynamics are equivalent to radial quantization.  This difference in the spectrum has a direct consequence on the behavior of unequal time two-point functions in the two phases. As was shown in \cite{Wen:2022pyj}, in the non-heating phase, the unequal two-point function shows an oscillatory behavior, while in the heating phase, it shows an exponential decay.  Thus in the heating phase, operators cluster in time and so in the formulation of \cite{Furuya:2023fei}, are `mixing' operators.  
\vskip 0.5cm

\noindent \underline{\large{\textbf{Dynamical phase classification via Von-Neumann Factors}}}\\
\vskip -0.2cm
\noindent We end this section by noting that the identification of the heating phase Hamiltonian with a Modular Hamiltonian of a subregion, provides an interesting non-equilibrium analogue of the {\it Hawking-Page} transition between the non-heating and heating phases in these systems \footnote{For a recent study see \cite{Miyata:2024gvr}.}. In \cite{Das:2022pez},  such a transition was also observed in the bulk dual in the first derivative of the free energy of a probe Brane, as one moved between the heating and non-heating phases. Here we point out an analogous transition directly in the CFT, in the language of Von-Neumann factors. 

On the one hand, the fact that the operators in the CFT are mixing operators under the heating phase Hamiltonian as well as the clustering property of (connected) two-point unequal time correlators under the flow generated by heating phase Hamiltonian in the vacuum state \footnote{The vacuum is a cyclic and separating state which serves the purpose of a TFD state in our context as we discussed in the introduction and hence satisfy KMS condition with respect to Heating phase Hamiltonian or the modular Hamiltonian.}\cite{Das:2022pez}, implies that one can associate a type $III_1$ algebra with the KMS state associated with the heating phase Hamiltonian\cite{Furuya:2023fei}.

On the other hand, the unitary equivalence of the non-heating phase Hamiltonian with the usual CFT Hamiltonian on a ring implies that at finite $c$, the thermal entropy is finite and so, one can associate a type-$I$ algebra {\it wrt} the KMS state (TFD state) associated with this Hamiltonian. 

Thus at finite $c$, one sees that as one changes the sign of $d$, one makes a transition from a type $I$ algebra to a type $III$ algebra. This is in contrast to the Hawking Page transition which happens strictly in the infinite $c$ limit, whereas one changes $\beta$, above a particular value of $\beta =\beta_{HP}$, the algebra associated with the TFD state transitions from type $I$ to type $III_1$ \cite{Leutheusser:2022bgi}. In the non-equilibrium case, the role of $\beta$ is played by $\delta$- the parameter that interpolates between the two phases.

\section{A view from the bulk}\label{bulk}

In this section, we will shift focus to the bulk description of these driven CFTs. In \cite{Das:2022pez}, the holographic dual-bulk geometries corresponding to the three distinct dynamics of the Floquet Hamiltonian $H_F$ in the boundary were discussed. The idea was to write down the bulk counterparts of the boundary Floquet Hamiltonians,  by replacing the global Virasoro generators $L_0, L_1, L_{-1}$ with their bulk representations and then solving the tangent curve generated by this bulk Hamiltonian.

The three phases of the $sl(2,\mathbb{R})$ drive correspond to the three distinct geometries, i.e.- $AdS_3$ metrics foliated by different $AdS_2$ slices. In particular, $AdS_2$ black hole, global $AdS$, and Poincar\'e patch correspond to the heating, non-heating, and phase transition respectively.

The curve equations in the bulk corresponding to the heating phase are given in \cite{Das:2022pez}\footnote{Here, we have changed the conventions of $x$ and $\tau$ compared to \cite{Das:2022pez} because we define the complex coordinate on the boundary as $\omega= \tau+ i x$ and $\Bar{\omega}=\tau-i x$. Also, we have changed the definition of $\delta$ to be $\delta=4 \beta \gamma-\alpha^2$.}

\begin{eqnarray}\label{bulkcurves}
    \tau &=& \frac{\sqrt{|\delta|}}{4 \beta}\left(\tan\left[\frac{\sqrt{|\delta|}}{2}(\sigma^0+i \theta)\right]+\tan\left[\frac{\sqrt{|\delta|}}{2}(\sigma^0-i \theta)\right]\right)\\\nonumber
    x &=& \frac{\sqrt{|\delta|}}{4 \beta i}\cos{\phi}\left(\tan\left[\frac{\sqrt{|\delta|}}{2}(\sigma^0+i \theta)\right]-\tan\left[\frac{\sqrt{|\delta|}}{2}(\sigma^0-i \theta)\right]\right)\\\nonumber
    u &=& \frac{\sqrt{|\delta|}}{4 \beta i}\sin{\phi}\left(\tan\left[\frac{\sqrt{|\delta|}}{2}(\sigma^0+i \theta)\right]-\tan\left[\frac{\sqrt{|\delta|}}{2}(\sigma^0-i \theta)\right]\right)
\end{eqnarray}

Putting these equations of the curves into the Poincaré metric yields a $AdS_2$ black hole foliation of $AdS_3$. Notice that if one takes the boundary limit, $\phi=0$, in \eqref{bulkcurves}, one recovers the conformal transformation in the CFT, between the conformal frames $CF_1$ and $CF_2$, discussed in equation \eqref{bdyconftran}, as one should. 

\subsection{AdS Rindler description of the heating phase}\label{eqv}


Given this boundary conformal transformation \eqref{bdyconftran}, there is a well-known way to lift them to bulk diffeomorphisms, thus obtaining a bulk metric. 
The bulk diffeomorphisms corresponding to boundary conformal transformations are given by \cite{Anand:2017dav}:

\begin{eqnarray}\label{kaplan}
    \omega &\rightarrow& f(z)-\frac{2 y^2 \left(f'(z)\right)^2 \Bar{f}''(\Bar{z})}{4 f'(z)\Bar{f}'(\Bar{z})+y^2 f''(z) \Bar{f}''(\Bar{z})}\\\nonumber
  \Bar{\omega} &\rightarrow& \Bar{f}(\Bar{z})-\frac{2 y^2 \left(\Bar{f}'(\Bar{z})\right)^2 f''(z)}{4 f'(z)\Bar{f}'(\Bar{z})+y^2 f''(z) \Bar{f}''(\Bar{z})}\\\nonumber
  u &\rightarrow& y \frac{4 \left(f'(z)\Bar{f}'(\Bar{z})\right)^{\frac{3}{2}}}{4 f'(z)\Bar{f}'(\Bar{z})+y^2 f''(z) \Bar{f}''(\Bar{z})}
\end{eqnarray}

And the transformed metric is:

\begin{equation}
    ds^2=\frac{dy^2+dz d\Bar{z}}{y^2}-\frac{1}{2}S\left(f,z\right)dz^2-\frac{1}{2}S\left(\Bar{f},\Bar{z}\right)d\Bar{z}^2+y^2 \frac{S\left(f,z\right) S\left(\Bar{f},\Bar{z}\right)}{4}dz d\Bar{z}
\end{equation}

where $S\left(f,z\right)\equiv \frac{f'''(z)f'(z)-\frac{3}{2}\left(f''(z)\right)^2}{\left(f'(z)\right)^2}$.

For the conformal transformations given in \eqref{bdyconftran}, the Schwarzians are:

\begin{eqnarray}
   S\left(f,z \right) &=&  \frac{|\delta|}{2} \\\nonumber 
   S\left(\Bar{f}, \Bar{z}\right) &=& \frac{|\delta|}{2}
\end{eqnarray}

Therefore, the metric in this case is:

\begin{equation}
    ds^2=\left(\frac{d y}{4}-\frac{1}{y}\right)^2 (d\sigma^{0})^{2}+\left(\frac{d y}{4}+\frac{1}{y}\right)^2 d\theta^{2}+\frac{dy^2}{y^2}
\end{equation}

If we transform the coordinate $y=\frac{2 \left(\sqrt{r^2-|\delta|}\ +\ r\right)}{|\delta|}$ and analytically continue $\sigma^0 \rightarrow i \sigma^0$, the metric becomes:

\begin{equation}\label{heatingmetric}
    ds^2= -(r^2-|\delta|)(d\sigma^0)^2 +\frac{1}{(r^2-|\delta|)} dr^2+r^2 d\theta^2
\end{equation}

This is precisely the three-dimensional AdS-Rindler metric. This is 'equivalent' to the bulk geometry corresponding to the heating phase, discussed in \cite{Das:2022pez}, which was an $AdS_2$ black hole foliation of $AdS_3$, in the sense that both describe the same bulk observers (since the time coordinate in both descriptions is the same) and are related by local diffeomorphisms. 

One can repeat this procedure for the case of the non-heating phase, the boundary conformal transformations for which are given by:

\begin{eqnarray}\label{bdyconftrannh}
    \omega &=& \tau+i x = -\frac{\sqrt{\delta}}{2 \beta}\coth\left[\frac {\sqrt{\delta}z}{2}\right]\\\nonumber
    \Bar{\omega} &=& \tau-i x = - \frac{\sqrt{\delta}}{2 \beta}\coth\left[\frac{ \sqrt{\delta}\Bar{z}}{2}\right]
\end{eqnarray}

Hence, in this case, the Schwarzians are $S\left(f,z\right)=S\left(\Bar{f},\Bar{z}\right)= -\frac{\delta}{2}$, and the metric is:

\begin{equation}
    ds^2=\left(\frac{d y}{4}+\frac{1}{y}\right)^2 (d\sigma^{0})^{2}+\left(\frac{d y}{4}-\frac{1}{y}\right)^2 d\theta^{2}+\frac{dy^2}{y^2}
\end{equation}

Following the coordinate redefinition $y=\frac{2 \left(\sqrt{\rho^2+\delta}+\rho\right)}{\delta}$ and analytical continuation of $\sigma^0 \rightarrow i \sigma^0$, the metric can be written as:

\begin{equation}\label{nheatingmetric}
    ds^2= -(\rho^2+\delta)(d\sigma^0)^2 +\frac{1}{(\rho^2+\delta)} d\rho^2+\rho^2 d\theta^2
\end{equation}

As we can notice, the above metric \eqref{nheatingmetric} corresponds to the global $AdS_3$ geometry. On the phase boundary, the boundary conformal transformations are simply:

\begin{eqnarray}\label{bdyconftranpb}
    \omega &=& \tau+i x = -\frac{1}{\beta z}\\\nonumber
    \Bar{\omega} &=& \tau-i x = -\frac{1}{\beta \Bar{z}}
\end{eqnarray}

Therefore, in this case, the Schwarzians of the transformations are zero. As a result, after the analytic continuation of $\sigma^0$, the metric is:

\begin{equation}\label{pbmetric}
    ds^2= \frac{-(d\sigma^0)^2+ d\theta^2+ dy^2}{y^2}
\end{equation}

The above metric describes the Poincaré patch of three-dimensional AdS spacetime.

\subsection{Fixed points as Ryu-Takayanagi surface in the bulk}\label{fprt}
In this section, we discuss the bulk meaning of the fixed points under the $sl(2,\mathbb{R})$ valued Floquet Hamiltonian flows in the boundary CFT.  
For the Hamiltonian, that we considered in \eqref{hdefor}, the fixed points are at
\begin{equation}\label{fx1}
    \mathcal{Z}_{f_{1,2}}= \frac{\alpha  \pm i \sqrt{|\delta|}}{2 \beta}
\end{equation}
In the bulk, the fixed points correspond to the set of points that do not flow in time. From the tangent equations of the curve generated by the Hamiltonian in the bulk ($u \neq 0$), we get \footnote{See appendix B of \cite{Das:2022pez} for more details.}
\begin{eqnarray}
    \frac{du(s)}{ds} &=& 2 \beta u \left(\tau- \frac{\alpha}{2 \beta}\right) \\
    \frac{dx(s)}{ds} &=& 2 \beta x \left(\tau- \frac{\alpha}{2 \beta}\right)  \\
    \frac{d\tau(s)}{ds} &=& \beta \left(\left(\tau -\frac{\alpha}{2 \beta}\right)^2-x^2-u^2- |\delta| \right)
\end{eqnarray}
Then, by definition, on the fixed points we will have $\frac{dz}{ds}=\frac{dx}{ds}=\frac{d\tau}{ds}=0$. Using these conditions, if we solve the above equations, we get the equations of the fixed points to be
\begin{eqnarray}
    \tau= \frac{\alpha}{2 \beta}\\\label{rt1}
    u^2+x^2=-\frac{|\delta|}{4 \beta^2} \label{rt}
\end{eqnarray}
There are a few important points to mention here. Firstly, while solving the equations we have assumed that, $z\ne 0$ which makes sense since we are solving the equations in bulk. Also, as we can see, \eqref{rt} has solutions only when $\delta \le 0$, i.e. in the heating phase and on the phase boundary. This simply means that there are no bulk-fixed points in the non-heating phase, and therefore a bulk observer can access the full spacetime and not just a sub-region like the entanglement wedge. \\

The above equation corresponds to the equation of the RT surface between two boundary fixed points. This can be verified very easily from the above equations. In the heating phase, when $\delta <0$, from \eqref{rt} we get $x= \pm \frac{\sqrt{|\delta|}}{2 \beta}$ on the boundary ($u=0$) and from the \eqref{rt1} we get $\tau= \frac{\alpha}{2 \beta}$. Therefore, the corresponding coordinates on the plane are, $ z \equiv \tau+i x=\frac{\alpha  \pm i \sqrt{|\delta|}}{2 \beta}$ which are precisely the fixed points \eqref{fx1}. 

\section{Virasoro algebra from the boundary and the bulk}

In section \ref{mhhp}, we argued that the heating-phase Hamiltonian acts as a modular Hamiltonian for a subregion between two fixed points. To study quantization under the heating-phase Hamiltonian, it is necessary to introduce an infinitesimally small ($\epsilon$) cut-off around the fixed points. This results in a discrete Virasoro algebra with a distinct central extension term. The algebra becomes continuous in the $\epsilon\rightarrow0$ limit. A cut-off in the boundary theory with conformal boundary conditions corresponds to a cut-off surface near the RT surface in the bulk. On the other hand, there has been a lot of discussion in the literature of the existence of  Virasoro algebra in the near horizon limit of any $D\ge 3$ dimensional black holes \cite{Solodukhin:1998tc, Carlip:2002be, Carlip:1999db, Carlip:2005zn, Guica:2008mu, Kang:2004js}. In this section, we will first review the heating phase quantization and the corresponding Virasoro algebra following the work of \cite{Tada:2019rls}. Then we will go through the general strategy as discussed in \cite{Solodukhin:1998tc}, where the near horizon Virasoro algebra was constructed. Notably, we observe that the central extensions in both these cases match exactly.

\subsection{A brief review of the Heating-Phase Quantization}\label{hpq}
To make this section self-consistent, we will go through the basic steps to the heating phase quantization discussed in \footnote{See \cite{Arzano:2024ogp} for a recent work on similar ideas.} \cite{Tada:2019rls}. 
Usually in 2D CFTs, one deals with the un-deformed Hamiltonian $H_{CFT}=L_0+\bar{L}_0-\frac{c}{12}$ which is the dilatation generator and generates time translation along the radial direction on the z-plane. This is known as the ``radial quantization" where the Hilbert space of a CFT defined on a constant (Euclidean) time slice on the cylinder $S^1\times R^1$ gets mapped to the Hilbert space on a constant radius surface on the plane $R^2$. 




In the Floquet CFT setup, one considers a more general effective Hamiltonian consisting of a linear combination of $L_0, L_1, L_{-1}$ and the corresponding conjugates.  

\begin{figure}
    \centering
    \includegraphics[scale=0.5]{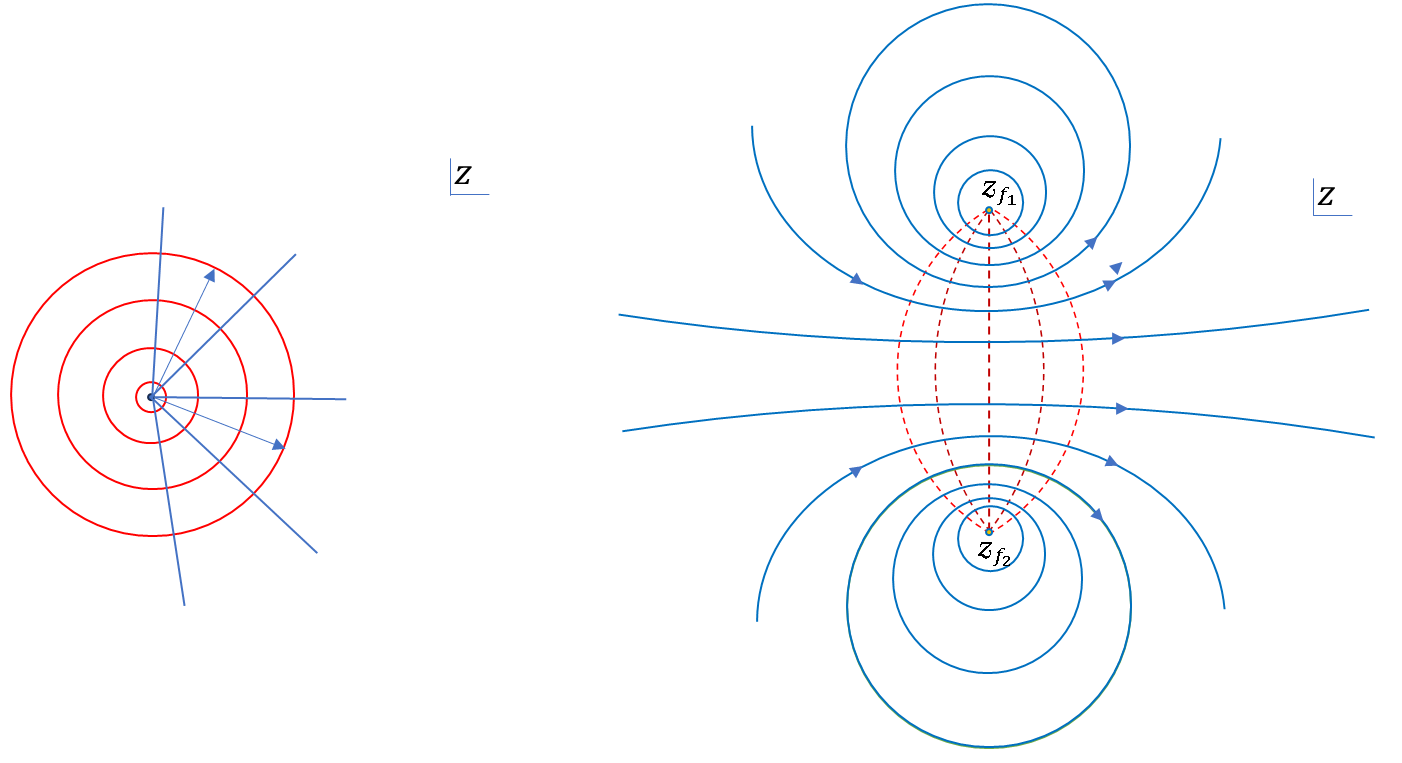}
    \caption{The figures illustrate two different quantization methods. The left-hand side figure represents radial quantization, while the right-hand side picture represents the Heating-phase quantization. In these figures, the red and blue colours indicate constant time and spatial slices, respectively, and the arrow represents the direction of time flow. In the second figure, the  points $z_{f_1}$ and $z_{f_2}$ represents the two fixed points of the underlying dynamics}
    \label{fig:enter-label}
\end{figure}

Given a heating phase Hamiltonian \eqref{hdefor} with $\delta < 0$, one can use the prescription in 
\cite{Ishibashi:2016bey,Tada:2019rls} to define the conserved charges corresponding to new time translation generated by \eqref{hdefor}, as the following 

\begin{equation}\label{hpqvg}
   \mathcal{L}_n=\frac{1}{2 \pi i}\int_{C} g(z) f_n(z) T(z) dz\quad \ and\quad    \mathcal{L}_0=\frac{1}{2 \pi i}\int_{C} g(z)  T(z) dz
\end{equation}
along with their respective conjugates, where $T(z)(\bar{T}(z))$ is the chiral (anti-chiral) component of the energy-momentum tensor) and the function $g(z)$ is obtained by rewriting the heating phase Hamiltonian in terms of the corresponding classical generators $l_0,l_1,l_{-1}$ and their differential forms $z\partial_z,z^2 \partial_z,\partial_z$. In terms of parameters of $H_{F}$, $g(z)=\beta z^2+ \alpha z + \gamma $ with $\delta=\alpha^2-4 \beta \gamma <0$. To be specific, $H_{F}= \mathcal{L}_{0}+\Bar{\mathcal{L}_{0}}$, whereas $\mathcal{L}_{n}$s are its eigenmodes with eigenvalue $n$. One must note that just like radial quantization, these conserved charges \eqref{hpqvg} are defined on a contour $C$ where the time generated by the floquet Hamiltonian $H_F$ in the heating phase remains constant. In radial quantization, one defines the usual Virasoro generators on the concentric circles of different radii at different times. In particular, $t=-\infty$ corresponds to the center, effectively giving rise to the well-known notion of state-operator correspondence. On the other hand, in this case of heating phase quantization, the time slices are arcs connecting two fixed points, and the constant spatial slices encircle the two fixed points with varying radii as well as varying centers (See Fig.\ref{fig:enter-label} for reference).
The $f_n(z)$ which appeared in the definition of $\mathcal{L}_{n}$ in \eqref{hpqvg}, is defined  as

\begin{equation}
    f_n(z)=e^{n \int g(z)dz}
\end{equation}
However, it is important to note that the function $f_n(z)$ depends crucially on the choice of $g(z)$. 
Therefore, unlike the usual Virasoro generators the label $n$ can take any value (continuous or discrete) depending on the final expression of $f_n(z)$ (for a given $g(z)$) and its single valued-ness on the complex plane. For the heating phase, \cite{Tada:2019rls} showed that the $f_n(z)$ has divergences at the fixed points of the dynamics. This makes the modes $\mathcal{L}_{n}$ to be non-analytic at the fixed points. In particular, to obtain the algebra of $\mathcal{L}_{n}$s, one needs to take care of these divergences. Therefore, one needs to introduce a cut-off $\epsilon$ circling the fixed points. Thus, introduction of cut-off will modify the constant-time contour $C \rightarrow C_{\epsilon}$ and the eigenmodes $\mathcal{L}_{n} \rightarrow \mathcal{L}_{n}^{\epsilon}$. Once we demand the $\mathcal{L}_{n}^{\epsilon}$s satisfy Virasoro algebra with finite central extension term \footnote{The corresponding classical space-time generators will satisfy the Witt algebra \cite{Ishibashi:2016bey}}, the generator labels $n$ in \eqref{hpqvg} are restricted to take discrete values, rendering the spectrum discrete \cite{Tada:2019rls} \footnote{See \cite{Das:2024mlx} for a review and the final form of the modified Virasoro algebra.}. In particular, $n$ takes an integer or half-integer value multiplied by a $\epsilon$ dependent fractional term. After a proper rescaling \cite{Tada:2019rls}, one would end up with the following form of the Virasoro algebra, where $n \in \mathbb{Z}$:


\begin{equation}\label{hpqva}
    [\mathcal{L}_{m}^{\epsilon},\mathcal{L}_{n}^{\epsilon}]=(m-n)\ \mathcal{L}_{n+m}^{\epsilon}+\frac{c}{12} \left( m^3+\rho^2 |\delta| m \right)\ \delta_{n+m,0}
    \end{equation}

Here, $\rho$ is a divergent quantity that depends on the system parameters: the distance between the two fixed points $L_{\delta}$, the Hamiltonian parameter $\beta$ and the cut-off $\epsilon$, and is given by $\rho=\frac{\log ( \frac{L_{\delta}}{\epsilon})}{\beta \pi L_{\delta}}$ and $L_{\delta}= \frac{\sqrt{|\delta|}}{\beta}$. 

It is noteworthy to mention that, due to the boundary condition placed at the cut-off, the whole problem is getting mapped to a BCFT on a finite cylinder as we mentioned earlier. As argued in \cite{Das:2024mlx}, the boundary conditions on those two cut-offs can be fixed by conformal boundary conditions. Hence, from the usual BCFT point of view, we will end up with a single copy of Virasoro algebra where the holomorphic and anti-holomorphic Virasoro generators would not be independent anymore. This makes \eqref{hpqva} to be a unique and distinct Virasoro algebra from the usual one.


\subsection{Bulk realization of the Virasoro algebra}\label{bulkvir}

The author argued in \cite{Solodukhin:1998tc} that the presence of the horizon leads to the breaking of one of the conformal symmetries, resulting in one copy of the Virasoro algebra near the horizon. To construct the conformal generators and the corresponding Virasoro algebra, one begins with the Einstein-Hilbert action and computes the action for the metric under consideration. In the case where $D \ge 3$, it was demonstrated that the action can be reformulated as the action of a 2D Liouville scalar field theory by integrating out the remaining compact directions. It was shown that in the near horizon limit, this scalar field theory becomes a 2D conformal field theory in flat space-time, and the generator of the conformal transformation can be expressed using the components of the energy-momentum tensor of the scalar field. For a metric like \eqref{heatingmetric}, one introduces a coordinate 
\begin{equation*}
    z=\int \frac{dr}{r^2- |\delta|}=\frac{1}{ 2 \sqrt{|\delta|}}\log{\left(\frac{r-\sqrt{|\delta|}}{r+\sqrt{|\delta|}}\right)}
\end{equation*}
in which the location of the horizon($r=\sqrt{|\delta|}$) is replaced by $z=-\infty$ and the boundary location ($r=\infty$) is replaced by $z=0$. In terms of this coordinate, the  near-horizon conformal generators are given by 
\begin{equation}\label{charge}
    T[\zeta]= \int_{-\tilde{l}}^{0}dz T_{++}(z) \zeta(z)
\end{equation}
where $T_{++}(z)$ denotes the right-moving energy-momentum tensor of the scalar field in the z coordinate, while $\zeta(z)$ represents the vector fields that generate diffeomorphisms preserving the horizon symmetry. Here we have introduced a cut-off $\tilde{l}$ in the $z$ coordinate, which we will take to infinity at the end of the computations. To be precise, $l$ is a near-horizon cut-off. In the usual AdS-Rindler coordinate \eqref{heatingmetric}, this amounts to take a cut-off at $r=\sqrt{|\delta|}+\zeta$. In terms of $\zeta$, the form of $l$ would be
\begin{align}\label{defn l}
    \tilde{l}= \frac{2}{\sqrt{|\delta|}}\log\left(\frac{2\sqrt{|\delta|}}{\zeta}\right).
\end{align}
Following this, we expand the vector fields into Fourier modes and introduce the bulk Virasoro generators.
\begin{equation}\label{bulk vir}
    L_{n}^{b}=\frac{l}{2 \pi} \int_{-\tilde{l}}^{0}dz T_{++}(z) e^{i \frac{2 \pi}{\tilde{l}} n z}
\end{equation}
As a result, the Poisson algebra of the charges \eqref{bulk vir} reproduces the Virasoro algebra as

\begin{equation}\label{nhva}
    i\{ L_{k}^{b}, L_{n}^{b} \}_{PB}=(k-n)\ L_{n+k}^{b}+ \frac{c}{12} k \left(k^2+ |\delta|\frac{\tilde{l}^2 }{4 \pi^2}\right)\ \delta_{n+k,0}
\end{equation}
In the above equation, the central charge depends on the classical value of the field on the horizon. One can immediately observe the resemblance between \eqref{nhva} and \eqref{hpqva} if $\tilde{l}= 2 \pi \rho$. Therefore, 

\begin{equation}\label{cut-off rel}
    \log\left({\frac{\sqrt{|\delta|}}{\beta \epsilon}}\right)= \frac{\sqrt{|\delta|}}{2} \tilde{l}
\end{equation}
In other words, from \eqref{defn l} we get $\zeta = 2\beta \epsilon$. Hence, the boundary cut-off $\epsilon$ is related to near-horizon cut-off.

In the next section, we will see that the same relation of the form \eqref{cut-off rel} between the boundary cut-off and the bulk IR (near boundary) cut-off would appear if one tries to match the entanglement entropy of the region between two fixed points with the entropy of the AdS-Rindler. This would suggest the near boundary bulk IR cut-off is proportionally related to the near horizon bulk UV cut-off. This fact could be well-explained if we study the same in the equivalent bulk picture in coordinates ($\phi,\sigma^{0},\theta$) described in \cite{Das:2022pez}. In that case, the compactified $\phi$ direction can be integrated out to carry out the similar analysis in the AdS$_{2}$ black hole patch in $(\sigma^{0},\theta)$ coordinates. In the boundary limit $(\phi \rightarrow 0,\pi)$, the horizon of the AdS$_{2}$ black hole exactly coincides with the fixed points, as we observed previously. Thus, the cut-off in $\theta$ is the same cut-off near the horizon of the 2D boundary black hole. The bulk description of the metric can be thought of as the pulling out of the boundary black hole in the bulk along the $\phi$ direction \footnote{At each constant $\phi$ slice, there is a 2D AdS black hole.}. The equivalence of this metric with the AdS-Rindler in the bulk would explain the equivalence of near-horizon and boundary cut-off as we observed.

\section{Entanglement Entropy as AdS-Rindler Entropy }\label{EE}

In this section, we will highlight that the entanglement entropy between the two fixed points is equal to the Rindler entropy corresponding to the AdS-Rindler metric. More specifically, we will compute the holographic Entanglement entropy by computing the length of the RT surface and comparing it with the Rindler entropy.\\

\noindent \underline{\large{\textbf{Entanglement Entropy}}}\\
\vskip -0.3cm
\noindent To compute the entanglement entropy using the Ryu-Takayanagi formula, one needs to find the length of the RT surface. For this, one computes the geodesic length between two arbitrary points ($u_1, t_1, x_1$) and ($u_2, t_2, x_2$), which is given by the following relation in AdS-Poincar\'e coordinates, 
\begin{equation}
    D= \cosh^{-1}\left( \frac{(x_1-x_2)^2-(t_1-t_2)^2+ u_1^2+u_2^2}{2 u_1 u_2}  \right)
\end{equation}

where, $x,t$ are the boundary coordinates while $u$ extends into the bulk. To avoid the divergence, one usually sets $u_1=u_2=\epsilon$. In our setup, $t_1=t_2=\frac{\alpha}{2 \beta}$, $x_1=\frac{\sqrt{|\delta|}}{2 \beta}$ and $x_2=-\frac{\sqrt{|\delta|}}{2 \beta}$. Then the RT formula tells us that the entanglement entropy is given by
\begin{equation}\label{entent}
    S_{ent}= \frac{D}{4 G_{\text{N}}^{(3)}}=\frac{1}{2 G_{\text{N}}^{(3)}}\log \left(\frac{\sqrt{|\delta|}}{\beta \epsilon}\right)=\frac{c}{3}\log \left(\frac{\sqrt{|\delta|}}{\beta \epsilon}\right)
\end{equation}
Note that, in the last equation, we have used  $G_{\text{N}}^{(3)}=\frac{3}{2 c}$.\\

\noindent \underline{\large{\textbf{Entropy of AdS-Rindler}}}\\
\vskip -0.3cm
\noindent The entropy of AdS-Rindler is defined as \cite{Compere:2018aar},
\begin{equation}
    S_{\text{H}}=\frac{1}{4 G_{\text{N}}^{(3)}}\int_{-\frac{1}{\epsilon'}}^{\frac{1}{\epsilon'}}d\theta \sqrt{g_{\text{ind}}}
\end{equation}
where $\sqrt{g_{\text{ind}}}$ is the determinant of the induced metric on the horizon. At a constant time $ds_{\text{ind}}^2= ds^2[r=\sqrt{|\delta|}, t=constant]=\delta (d\theta)^2$, so
\begin{equation}\label{entrind}
    S_{\text{H}}=\frac{\sqrt{|\delta}|}{2 G_{\text{N} }\epsilon'}
\end{equation}
Therefore, \eqref{entrind} and \eqref{entrind} matches if we identify $\epsilon'$ and $\epsilon$ as
\begin{equation}\label{epsilon}
\frac{\sqrt{|\delta|}}{\epsilon'}=\log\left(\frac{\sqrt{|\delta|}}{\beta \epsilon}\right)
\end{equation}
This can be understood from the relation between the coordinate u and $\theta$, which we know using \eqref{bdyconftran} in \eqref{kaplan} to be

\begin{equation*}
    u=\left(\frac{\sqrt{|\delta|}}{2 \beta}\right)\frac{(\cosh{(\sqrt{|\delta|}\sigma^0)}+\cosh{(\sqrt{|\delta|}\theta)})}{(\cosh{(\sqrt{|\delta|}\sigma^0)}+\cosh{(\sqrt{|\delta|}\theta)})\cosh{(\sqrt{|\delta|}\theta})}
\end{equation*}
In the large $\theta$ limit this reduces to,
\begin{equation}
    u=\frac{\sqrt{|\delta|}}{\beta} e^{-\sqrt{|\delta|}\theta}
\end{equation}
If we put $u=\epsilon$ and $\theta=\frac{1}{\epsilon'}$, it precisely reproduces \eqref{epsilon}.\\

\section{Discussion}\label{disc}

In this article, we have explored some consequences that follow from the identification of the heating phase Hamiltonian with the Modular Hamiltonian of a sub-region, in the vacuum state. From this perspective, the heating phase quantization describes the evolution of operators inside the causal diamond of the subregion, between two fixed points of the $sl(2,\mathbb{R})$ flow, in modular/stroboscopic time. To study the {\it modular quantization} of this subregion, one needs to transform to a conformal frame, wherein the time coordinate is identified with the modular time. One then imposes a conformal boundary condition at an $\epsilon$ cut-off away from the fixed points of the Floquet Hamiltonian flows. This reduces the system of study to that of a BCFT on a strip. 
We also showed that holographically, this BCFT describes the dynamics inside the RT Entanglement Wedge, described by an AdS Rindler metric, where the Rindler Horizon is identified with the RT surface, and that the boundary cut-off, near the fixed points, translates to a cut-off, analogous to a stretched horizon\footnote{See \cite{Das:2024mlx} and references therein where the connection to the stretched horizon has been explored in details.}, near the RT surface or the Rindler horizon. We further argued that the single copy of the conformal algebra preserved by the boundary conditions of the BCFT should be dual to the single copy of the Virasoro symmetry at the stretched Horizon. From the perspective of driven CFTs, we also argued that the different dynamical phases of these $CFT$s can be classified by the type of Von Neumann factors. In particular as one moves from the non-heating to the heating phase, by tuning the external control parameters like the frequency or time period of the drive, the factor changes from type $I$ to type $III$. This, again, is a direct consequence of the identification of the heating phase dynamics with Modular quantization.  

Since in the bulk description, the qualitative differences in the three dynamical phases correspond to differences in observations made by three different classes of observers \cite{Das:2022pez}, it follows that the Von-Neumann algebra of observables accessible to the observer, dual to the heating phase would be type $III$, while the corresponding Von-Neumann algebra of observables accessible to the observer dual to the non-heating phase should be Type $I$. It would be interesting to understand this better along the lines of recent investigations \cite{Witten:2023qsv, Penington:2023dql, Strohmaier:2023opz, Kudler-Flam:2024psh, Chen:2024rpx}.

Given that the boundary theory in our work describes the AdS Rindler Black hole, it should be possible to explicitly construct the microstates of the black hole directly from the heating phase of the driven CFT. In fact, these geometries would be the so-called fixed area states, since these would be eigenstates of the $H^{(u)}_{B_1 B_2}$, which in the large $c$ is identified with the area operator\cite{Dong:2018seb}. It would be nice to explicitly construct these states directly in the CFT\footnote{Partial progress along this line has been done in \cite{Das:2024mlx}.}. 

It would also be interesting to understand better the connection between the `emergent' horizon Virasoro algebra and the `emergent' boundary Virasoro algebra in Modular quantization. In fact, as shown in \cite{Das:2024mlx}, the microcanonical counting of states of this Hamiltonian which also describes the horizon CFT, would exactly reproduce the Bekenstein-Hawking entropy of BTZ. In the context of horizon CFT, this implies that these states live in the (vicinity of) horizon rather than at an asymptotic boundary. Also, a similar structure of near horizon Virasoro algebra exists in any higher dimensional asymptotic AdS or flat spacetime, as argued in \cite{Solodukhin:1998tc}. Hence, we suspect the appearance of this conformal field theory and the emergence of horizon Virasoro symmetry have far-reaching consequences, beyond AdS/CFT.  We hope to explore some of these issues in the immediate future.    


\vskip 0.4cm
	\section*{Acknowledgments}
	
	The authors are thankful to Souvik Banerjee, Parthajit Biswas, Diptarka Das, Alok Laddha, and Partha Paul for several useful discussions. The authors thank Arnab Kundu and Krishnendu Sengupta for their helpful discussions on related topics. BR would like to thank Aninda Sinha for the three-week visit to IISc, where, some final works of this project were completed. BE acknowledges the support provided by the SERB grant CRG/2021/004539. The research work of SD is supported by a DST Inspire Faculty Fellowship. The work of SP and BR is supported by the Senior Research Fellowship(SRF) funded by the University Grant Commission(UGC) under the CSIR-UGC NET Fellowship.

\appendix
\section{Relation among different Non-heating phase Hamiltonians}\label{apnA}

As discussed earlier in section \ref{mhhp}, it was stated that all the Hamiltonians in the non-heating phase are unitarily equivalent. This section aims to furnish the necessary mathematical details to substantiate this claim. Note that, in what follows, we will only discuss using the holomorphic part of the theory, which remains consistent for the anti-holomorphic counterparts.
The first key point to keep in mind is that the subalgebra of the Virasoro algebra,

\begin{equation}
    [J_0,J_{\pm}]=\pm J_{\pm }, [J_{-},J_{+}]=2 J_0
\end{equation} 
obtained by redefining $L_{\pm n},L_0$ as 
$$J_{\pm}=\frac{L_{\mp n}}{n}\, \ \ J_0=\frac{L_0}{n}
+\frac{c}{24 n} (n^2-1)$$

enables to show that the following condition is true 


\begin{equation}\label{a1c1}
    e^{\alpha J_0+\beta J_+ + \gamma J_-}= e^{b J_+} e^{a J_0} e^{ c J_-}
\end{equation}
where the coefficients $\alpha,\beta,\gamma$ and $a,b,c$ are related by the following relation as discussed in \cite{Liska:2022vrd, Matone:2015wxa},

\begin{eqnarray}\label{a1c2}
a=Exp\bigg[\frac{1}{(\cosh \mathcal{P}-\frac{\alpha}{2 \mathcal{P}}\sinh \mathcal{P})^2}\bigg], \nonumber\\
    b= \frac{\frac{\beta}{\mathcal{P}} \sinh \mathcal{P}}{\cosh \mathcal{P}-\frac{\alpha}{2 \mathcal{P}}\sinh \mathcal{P}},\ c= \frac{\frac{\gamma}{\mathcal{P}} \sinh \mathcal{P}}{\cosh \mathcal{P}-\frac{\alpha}{2 \mathcal{P}}\sinh \mathcal{P}}  
\end{eqnarray}
where,
\begin{equation}
    \mathcal{P}=\frac{\sqrt{\alpha^2-4 \beta \gamma}}{2}
\end{equation}
The condition in \eqref{a1c1} then allows writing the following similarity transformation ($\mathcal{S}=e^{ (\alpha J_0+\beta J_+ + \gamma J_-)}$) acting on $J_0$ as,
\begin{equation}\label{a1c3}
 \mathcal{S} J_0 \mathcal{S}^{-1} =  e^{ b J_+} e^{a J_0} e^{ c J_-} J_0  e^{-c J_-} e^{-a J_0} e^{- b J_+}
\end{equation}

This crucial step makes it all easy and enables us to utilize the Baker-Campbell-Hausdorff formula and exploit the sub-algebra to carry out the computation step by step to find,

\begin{equation}
  e^{c J_-} J_0 e^{-c J_-} =1+ c[J_-,J_0]+ \frac{c^2}{2!}[j_-,[J_-,[J_-,J_0]]+\dots= J_0+c J_-    
\end{equation}

Similarly ,
\begin{equation}
   e^{a J_0} e^{ c J_-} J_0  e^{-c J_-} e^{-a J_0}=  e^{a J_0} (J_0+c J_-)e^{-a J_0}= J_0+(c e^{-a}) J_-
\end{equation}
and 
\begin{equation}
   e^{ b J_+} e^{a J_0} e^{ c J_-} J_0  e^{-c J_-} e^{-a J_0} e^{- b J_+}=  (1- 2 b c e^{-a} )J_0+ (c b^2 e^{-a}-b) J_+ + (c  e^{-a}) J_-
\end{equation}
At this stage, one should check that while this similarity transformation takes $J_0\rightarrow (p J_0 + q J_+ + r J_-)$, the Casimir $p^2-4 q r$ remains the same, specifically, $p^2-4 q r=1$ for the above case. From this one can conclude that the usual CFT hamiltonian in radial quantization $L_0+\bar{L}_0$, which is an example of a non-heating phase hamiltonian gets similarity transformed to another non-heating phase hamiltonian by retaining the Casimir invariant.

Using a very similar approach, one can also show that a unitary transformation of $L_0$ with the floquet generator $ U_F=e^{-i H_F s}$, where, $H_F=\alpha L_0+\beta L_+ + \gamma L_-$ takes $L_0$ to some other linear combination of $L_0$ retaining the casimir invariant. Note that, here $$L_+=\frac{L_1+L_{-1}}{2}, L_-=\frac{L_1-L_{-1}}{2 i}$$ and $\alpha,\beta,\gamma$ are real. Following the previous discussion, one begins with the equivalence,
\begin{equation}\label{a2c1}
    e^{i H_F s} L_0 e^{-i H_F s}=  (e^{i k L_-} e^{ i g L_+} e^{i f L_0}) L_0  (e^{-i f L_0} e^{-i g L_+} e^{-i k L_-} )
\end{equation}
where a relation similar to that of \eqref{a1c2} exists among the coefficients $(\alpha,\beta,\gamma)$ and $(f,g,k)$. Using \eqref{a2c1} and exploiting the BCH formula and 

\begin{equation}
    [L_{\pm},L_0]=\pm i L_{\mp}, \ \ \ \  [L_+,L_-]=i L_0
\end{equation}
one can find that, 

\begin{eqnarray}
    e^{i g L_{\pm}} L_0 e^{- i g L_{\pm}}&=& (\cosh g)L_0\mp(\sinh g)L_{\mp}  \\
\end{eqnarray}

  and therefore, the RHS of \eqref{a2c1} to be  $(\cosh g \cosh k )L_0+ (\cosh g\cosh k)L_+- \sinh g$. Note that the Casimir ($\alpha^2-\beta^2-\gamma^2$) remains invariant ($=1$) for this too.

\section{Fixed points as the end points of subregion}\label{apnB}

In this section, we want to show explicitly that the Hamiltonian in the heating phase corresponds to the modular Hamiltonian of a sub-region, with the endpoints being the fixed points of the conformal transformation generated by the Hamiltonian. 
The Hamiltonian in consideration is, 
\begin{equation}\label{hdef1}
    H_{F}= (\alpha L_0 +\beta L_{1}+\gamma L_{-1})+ c.c -\frac{c}{12}
\end{equation}

As the above Hamiltonian is made out of only the global generators, it generates M\"{o}bius transformation ($z'=\frac{a z+b}{c z+d}$) in the complex plane. In the heating phase, when $(\alpha^2-4 \beta \gamma)<0$, the parameters of the transformation generated by \eqref{hdef1} are 
\begin{align}\label{confo}
    a &= -\cos{\frac{|\delta| \pi \tau}{L}}-\frac{\alpha}{|\delta|} \sin{\frac{|\delta| \pi \tau}{L}}\nonumber\\
    b &= -\frac{\beta+ i \gamma}{|\delta|} \sin{\frac{|\delta| \pi \tau}{L}}\nonumber\\
    c &= -\frac{\beta - i \gamma}{|\delta|} \sin{\frac{|\delta| \pi \tau}{L}}\nonumber\\
    d &= -\cos{\frac{|\delta| \pi \tau}{L}}+\frac{\alpha}{|\delta|} \sin{\frac{|\delta| \pi \tau}{L}}
\end{align}
Now, for a conformal transformation $z \rightarrow z'=\frac{a z+b}{c z+d}$, the fixed points are 
\begin{equation}\label{fixed}
    \mathcal{Z}_{f_{1,2}}=\frac{a-d \pm \sqrt{(a-d)^2+4 b c}}{2 c}
\end{equation}
We put \eqref{modular11}, \eqref{modular12}, \eqref{modular13} in \eqref{confo} to write the parameters of the conformal transformation in terms of $R_1$ and $R_2$ and then we use \eqref{fixed} to get 
\begin{align}
    \mathcal{Z}_{f_{1}} &= e^{\frac{2 \pi i R_2}{L}}\nonumber\\
    \mathcal{Z}_{f_{2}} &= e^{\frac{2 \pi i R_1}{L}}
\end{align} Therefore, we can conclude that two endpoints (on the cylinder) of the region corresponding to the Modular Hamiltonian correspond to the two fixed points.

		\end{document}